\documentclass[prb,twocolumn,showpacs,preprintnumbers,amsmath,amssymb,floatfix]{revtex4}
\usepackage{graphicx}
\usepackage{dcolumn}
\usepackage{revsymb}
\usepackage{longtable}
\usepackage{subfigure}
\usepackage{psfrag}                 
\newcommand{\figurewidth}{\columnwidth}
\begin{document}
\title{Many-Body perturbation theory calculations on circular quantum dots}
\author{E. Waltersson}
\author{E. Lindroth}
\affiliation{Atomic Physics, Fysikum,
Stockholm University, S-106 91 Stockholm, Sweden}

\date{\today}

\begin{abstract}
The possibility to use perturbation theory to systematically improve
calculations on circular quantum dots is
investigated.  A few different starting points, including Hartree-Fock,
are tested and the importance of correlation
is discussed. Quantum dots with up to twelve electrons  are treated and the effects
of an external magnetic field are examined.
The sums over excited states are carried out with a complete
finite radial basis set obtained through the use of B-splines.
The calculated addition energy spectra are
compared with experiments and the implications for the filling sequence of
the third shell are discussed in detail. 
\end{abstract}


\maketitle

\section{Introduction\label{sec:intro}}

During the last decade a new field on the border between condensed
matter physics and atomic physics has emerged. Modern semi-conductor
techniques allow fabrication of electron quantum confinement devices,
called quantum dots, containing only a small and controllable number
of electrons. The experimental techniques are so refined that one
electron at a time can be injected into the dot in a fully
controllable way. This procedure has shown many similarities between
quantum dots and atoms, for example the existence of shell structure.
To emphasize these similarities quantum dots are often called
artificial atoms. The interest in quantum dots is mainly motivated by
the fact that their properties are tunable through electrostatic gates
and external electric and magnetic fields, making these {\em designer
atoms} promising candidates for nanotechnological applications. An
additional aspect is that quantum dots provide a new type of targets
for many-body methods. In contrast to atoms they are essentially
two-dimensional and their physical size is several orders of magnitude
larger than that of atoms, leading e.g. to a much greater sensitivity
to magnetic fields. Another difference compared to atoms is the
strength of the overall confinement potential relative to that of the
electron-electron interaction, which here varies over a much wider
range.

The full many-body problem of quantum dots is truly complex. A dot is
formed when a two-dimensional electron gas in an heterostructure layer
interface is confined also in the $xy$--plane. The, for this purpose
applied gate voltage, results in a potential well, the form of which is
not known. A quantitative account of this trapping potential is one of
the quantum dot many-body problems. Self-consistent solutions of the
combined Hartree and Poisson equations by Kumar {\it et
al.}~\cite{kumar90} in the early nineties indicated that for small
particle numbers this confining potential is parabolic in shape at
least to a first approximation. Since then a two-dimensional harmonic
oscillator potential have been the standard choice for studies
concentrating on the second many-body problem of quantum dots; that of
the description of the interaction among the confined electrons. The
efforts to give a realistic description of the full physical
situation, see
e.g.~\cite{kumar90,jovanovic94,MatagneRealisticVQD,matagne02,melnikov05}
have, however, underlined that it is important to realize the limits
of this choice. To start with the pure parabolic potential seems to be
considerably less adequate when the number of electrons is
approaching twenty. The potential strength is further not independent
of the number of electrons put into the dot, an effect which is
sometimes approximately accounted for by decreasing the strength with
the inverse square root of the number of electrons~\cite{koskinen97}.
Finally, the assumption that the confining potential is truly
two-dimensional is certainly an approximation and it will to some
extent exaggerate the Coulomb repulsion between the electrons. In
Ref.~\cite{MatagneRealisticVQD} the deviation from the pure
two-dimensional situation is shown to effectively take the form of an
extra potential term scaling with the fourth power of the distance to
the center and which can be both positive and negative. Although the
deviation is quite small it is found that predictions concerning the
so called third shell can be affected by it.

There is thus a number of uncertainties in the description of
quantum dots. On the one hand there is the degree to which real dots
deviate from two-dimensionality and pure parabolic confinement. On
the other hand there is the uncertainty in the account of electron
correlation among the confined electrons. The possible interplay
among these uncertainties is  also an open question. In a situation
like this it is often an advantage to study one problem at a time,
since it is then possible to have  control over the approximations
made and quantify their effects. We concentrate here on the problem
of dot-electron correlation. For this we employ  a model dot; truly
two-dimensional, with perfect circular symmetry and with a well
defined strength of the confining potential. This choice is
sufficient when the aim is to test the effects of the approximations
introduced through the approximative treatment of the
electron-electron interaction.

Especially the experimental study by Tarucha et. al. \cite{Tarucha}
has worked as a catalyst for a vast number of theoretical studies of
quantum dots. A review of the theoretical efforts until a few years
ago has been given by Reimann and Manninen~\cite{reimannRMP02}. A
large number of calculations has been done within the framework of
Density Functional Theory
(DFT)~\cite{koskinen97,macucci97,lee98,reimannRMP02} and reference
therein, but also Hartree--Fock
(HF)~\cite{fujitoHF,szafran99,yannouleas99}, 
Quantum Monte Carlo methods~\cite{Ghosal,Saarikoski} and Configuration Interaction
(CI)~\cite{ReimannCI, Bruce,Szafran} studies have been performed. The
DFT--studies have been very successful. The method obviously accounts
for a substantial part of the electron-electron interaction. Still,
the situation is not completely satisfactory since there is no
possibility to systematically improve the calculations or to estimate
the size of neglected effects. For just a few electrons the
CI-approach can produce virtually exact results, provided of course
that the basis set describes the physical space well enough. The size
of the full CI problem grows, however, very fast with the number of
electrons and to the best of our knowledge the largest number of
electrons in a quantum dot studied with CI is six. It would be an
advantage to also have access to a many-body method which introduces
only well defined approximations and which allows a quantitative
estimate of neglected contributions. The long tradition of accurate
calculations in atomic physics has shown that Many-Body Perturbation
Theory (MBPT) has these properties. It is an in principle exact
method, applicable to any number of electrons, and the introduced
approximations are precisely defined. With MBPT it is possible to
start from a good, or even reasonable, description of the artificial
atom and then refine this starting point in a controlled way. We are
only aware of one study on quantum dots that have been done with MBPT,
the one by Sloggett and Sushkov\cite{Slogget}. They did second--order
correlation calculations on circular and elliptical dots in an
environment free of external fields.

In the present study we use second--order perturbation theory to
calculate energy spectra for quantum dots with and without external
magnetic fields.  We consider this second--order treatment as a
first step towards the calculation of correlation to high orders
through iterative procedures, an approach  commonly used for
atoms~\cite{mbpt}. The method is described in
Section~\ref{sec:method}. In section ~\ref{sec:validity} we compare
our calculations with experimental
results\cite{Tarucha,MatagneRealisticVQD},
DFT--calculations~\cite{ReimannCI} and CI--calculations, our own as
well as those of Reimann {\it et al.}~\cite{ReimannCI} and discuss
the strength and limits of the MBPT approach. We have mainly used
the Hartree-Fock description as starting point for the
perturbation expansion, but we also show examples with a few
alternative starting points, among them DFT. To obtain a complete
and finite basis set, well suited to carry out the perturbation
expansion, we use so called B-splines, see e.g. Ref.~\cite{deboor}.
The use of B-splines in atomic physics was pioneered by Johnson and
Sapirstein~\cite{johnson:86:spline-letter} twenty years ago and
later it has been the method of choice in a large number of studies
as reviewed e.g. in Ref.~\cite{bachau:01}. We compare our correlated
results to our own HF--calculations, thereby highlighting the
importance of correlation both when the quantum dot is influenced by
an external magnetic field and when it is not. We present addition
energy spectra for the first twelve electrons. The interesting third
shell (electron seven to twelve) is discussed in
Section~\ref{sec:result}. Here we investigate several different
filling sequences and show that correlation effects in many cases can
change the order of which the shells are filled. We note also that
the energy of the first excited state can be very close to the
ground state, in some case less than 0.1 meV, which raises the
question if it is always the ground state which is filled when an
additional electron is injected in the dot since more than one state
may lie in the transport window controlled by the source drain
voltage~\cite{Kouwenhoven}.

\section{Method\label{sec:method}}

The essential point in  perturbation theory is to divide the full
Hamiltonian $\hat{H}$ into a first approximation, $\hat{h}$,
and a correction, $\hat{U}$.
The first approximation should be easily
obtainable, in practice it is more or less always chosen to be an effective
one-particle Hamiltonian, and it should describe the system well enough to
ensure fast and steady convergence of the perturbation expansion. The partition is
written as
\begin{equation}
\label{eq:perturb}
\hat{H} = \sum_{i=1}^N \hat{h}(i) + \hat{U}.
\end{equation}
Here we have chosen to mainly use the Hartree-Fock Hamiltonian as
$\hat{h}$ but we have also investigated the possibility to use a few
other starting points.

A first approximation to the energy is obtained from the expectation
value of $\hat{H}$, calculated with a wave function in the form of a
Slater determinant formed from eigenstates to $\hat{h}(i)$. The result
is then subsequently refined through the perturbation expansion. Below
we describe the calculations step by step.

\subsection{Single-particle treatment}
For a single particle confined in a circular quantum dot the Hamiltonian reads
\begin{equation}
\hat{h}_s=\frac{\mathbf{\hat{p}}^2}{2m^*}+\frac{1}{2}m^*\omega^2r^2+\frac{e^2}{8m^*}B^2r^2+\frac{e}{2m^*}B\hat{L}_z+g^*\mu_b
B\hat{S}_z,
\end{equation}
where B is an external magnetic field applied perpendicular to the dot. The
effective electron mass is denoted with $m^*$ and the effective g-factor with $g^*$.
Throughout this work we use  $m^*=0.067m_e$ and $g^*=-0.44$, corresponding to bulk values in
GaAs.

The single particle wave functions separate in
polar coordinates as
\begin{equation}
\label{wavefunctionexpansion_eq}
|\Psi_{nm_{\ell}m_s}\rangle = |u_{nm_{\ell}m_s}(r)\rangle|e^{im_{\ell}\phi}\rangle|m_s\rangle.
\end{equation}

As discussed in the introduction  we expand the radial part of our wave functions in so
called B-splines labeled $B_i$ with coefficients $c_i$, i.e.
\begin{equation}
| u_{nm_{\ell}m_s}(r) \rangle = \sum_{i=1} c_i |B_i(r)\rangle.
\label{Bspline_expansion_eq}
\end{equation}
B--splines are piecewise polynomials of a chosen order $k$, defined on
a so called knot sequence and they form a complete set in the space
defined by the knot sequence and the polynomial order~\cite{deboor}.
Here we have typically used $40$ points in the knot sequence,
distributed linearly in the inner region and then exponentially
further out. The last knot, defining the box to which we limit our
problem is around 400 nm. The polynomial order is six and combined
with the knot sequence this yield $33$ radial basis functions,
$u_{nm_{\ell}m_s}(r)$, for each combination $(m_{\ell},m_s)$. The lower energy
basis functions are physical states, while the higher ones are
determined mainly by the box. The unphysical higher energy states are,
however, still essential for the completeness of the basis set.

Eqs. (\ref{wavefunctionexpansion_eq}) and (\ref{Bspline_expansion_eq})
imply that the Schr\"odinger equation can be
written as a matrix equation
\begin{equation}
\mathbf{Hc}=\epsilon\mathbf{Bc}
\label{Matrix_eq}
\end{equation}
where $H_{ji}=\langle B_j
e^{im\theta}|\hat{H}|B_ie^{im\theta}\rangle$ and
$B_{ji}=\langle B_j|B_i \rangle$
\footnote{Note that $\langle B_j|B_i \rangle\neq\delta_{ji}$ in general since B--splines of order
larger than one are non--orthogonal.}.

Eq.( \ref{Matrix_eq}) is a generalized eigenvalue problem that can be
solved with standard numerical routines. The integrals in
(\ref{Matrix_eq}) are calculated with Gaussian quadrature and since
B-splines are piecewise polynomials this implies that essentially no
numerical error is produced in the integration.

\subsection{Many-Body treatment}

The next step is to allow for several electrons in the dot and
then to account for the electron-electron interaction,
\begin{equation}
\frac{e^2}{4\pi\epsilon_r\epsilon_0}\frac{1}{\mid \mathbf{r}_i-\mathbf{r}_j \mid},
\end{equation}
where $\epsilon_r$ is the relative dielectric constant which in the
following calculations is taken to be $\epsilon_r=12.4$ corresponding
to the bulk value in GaAs. For future convenience we define the
electron--electron interaction matrix element as
\begin{equation}
\label{ee-interaction_eq}
\langle ab| \frac{1}{\hat{r}_{ij}}| cd \rangle =
\int\!\!\!\int
\frac{e^2\Psi_a^*(\mathbf{r}_i)\Psi_b^*(\mathbf{r}_j)\Psi_c(\mathbf{r}_i) \Psi_d(\mathbf{r}_j)}
{4\pi\epsilon_r\epsilon_0|\mathbf{r}_i-\mathbf{r}_j|}
dA_{i}dA_{j},
\end{equation}
where $a,b,c$ and $d$ each denote a single quantum
state i.e. $|a\rangle=|n^a,m^a_{\ell},m^a_s\rangle$.

\subsubsection{The Multipole expansion}
 As suggested by Cohl et. al\cite{Cohl}, the inverse radial distance
 can be expanded in cylindrical coordinates $(R,\phi,z)$ as
\begin{equation}
\label{multipole_eq}
\frac{1}{|\mathbf{r}_1-\mathbf{r}_2|} =
\frac{1}{\pi\sqrt{R_1R_2}}\sum_{m=-\infty}^{\infty}Q_{m-\frac{1}{2}}(\chi)e^{im(\phi_1-\phi_2)},
\end{equation}
where
\begin{equation}
\label{chi_eq}
\chi = \frac{R_1^2+R_2^2+(z_1-z_2)^2}{2R_1R_2}.
\end{equation}
Assuming  a 2D confinement we set  $z_1=z_2$ in (\ref{chi_eq}).
The $Q_{m-\frac{1}{2}}(\chi)$--functions are Legendre functions of the
second kind and half--integer degree. We evaluate them using a
modified\footnote{It is modified in the sense that we have changed the
  limit of how close to one the argument $\chi$ can be. This is
simply so that we can get sufficient numerical precision.} version
of software \textsf{DTORH1.f} described in\cite{Segura}.

Using (\ref{multipole_eq}) and (\ref{wavefunctionexpansion_eq}) the
electron--electron interaction matrix element
(\ref{ee-interaction_eq}) becomes
\begin{eqnarray}
\label{ee_interaction_BIG_eq}
\langle ab| \frac{1}{\hat{r}_{12}}| cd \rangle = \frac{e^2}{4\pi\epsilon_r\epsilon_0}
\langle u_{a}(r_i)u_{b}(r_j)|
\frac{Q_{m-\frac{1}{2}}(\chi)}{\pi\sqrt{r_ir_j}}
| u_{c}(r_i)u_{d}(r_j) \rangle \nonumber \\
\times
\langle e^{im_a\phi_i}e^{im_b\phi_j}|
\sum_{m=-\infty}^{\infty} e^{im(\phi_i-\phi_j)}
|e^{im_c\phi_i}e^{im_d\phi_j}\rangle \nonumber \\
\times
\langle m_s^a|m_s^c\rangle
\langle m_s^b|m_s^d\rangle.
\end{eqnarray}

Note that the angular part of (\ref{ee_interaction_BIG_eq}) equals
zero except if $m=m_a-m_c$ or $m=m_d-m_b$. This is how the degree of
the Legendre--function in the radial part of
(\ref{ee_interaction_BIG_eq}) is chosen. It is also clear from
(\ref{ee_interaction_BIG_eq}) that the electron--electron matrix element
(\ref{ee-interaction_eq}) equals zero if state $a$ and $c$ or state
$b$ and $d$ have different spin directions.

\subsubsection{Hartree--Fock}

If the wave function is restricted to be in the form of a single Slater
determinant,  the Hartree-Fock approximation can be shown to yield the lowest energy.
In this approximation 
each electron is governed by the
confining potential and an average {\em Hartree-Fock potential}
formed by the other electrons. To account for the latter  the
Hamiltonian matrix $\mathbf{H}$ in Eq. (\ref{Matrix_eq}) is modified
by the addition of a term:
\begin{equation}
u^{HF}_{ji}=\langle B_j|\hat{u}_{HF}|B_i \rangle=
\sum_{a\le N} \langle B_j a|\frac{1}{\hat{r}_{12}}|B_i a\rangle-\langle B_ja|\frac{1}{\hat{r}_{12}}|aB_i\rangle.
\label{HFoperator}
\end{equation}
 The sum here runs over all occupied orbitals, $a$, defined by quantum numbers $n$, $m_{\ell}$, and $m_s$.
Eq. (\ref{Matrix_eq}) is then solved iteratively yielding new and better
wave functions in each iteration until the energies become
self--consistent. The hereby obtained solution is often labeled the
unrestricted Hartree-Fock approximation since no extra constraints are imposed on $u^{HF}$.

One property of the unrestricted Hartree--Fock approximation deserves special attention.
Consider the effects of the Hartree-Fock potential on an electron in orbital
$b$,
\begin{equation}
\langle b |\hat{u}_{HF}|  b\rangle = \sum_{a\le N} \langle b a|\frac{1}{\hat{r}_{12}}|b
a\rangle-\langle b a|\frac{1}{\hat{r}_{12}}|a b\rangle,
\label{HForbital}
\end{equation}
where the last term in Eq.~(\ref{HForbital}), the exchange term, is
non-zero only if orbital $a$ and $b$ have the same spin. For
configurations where not all electron spins are paired electrons with
the same quantum numbers $n$ and $m_{\ell}$, but with different spin
directions, will experience different potentials. This is in
accordance with the physical situation, but has also an undesired
consequence; the total spin, ${\bf S}^2 = \left( \sum_i {\bf s}_i
\right)^2$, does not commute with the Hartree-Fock Hamiltonian. This
means that the state vector constructed as a single Slater determinant
of Hartree--Fock orbitals will not generally be a spin eigenstate.
However, the full Hamiltonian, Eq.~\ref{eq:perturb}, still commutes
with ${\bf S}^2$ and during the perturbation expansion the spin will
eventually be restored, provided of course that the perturbation
expansion converges. Since, in contrast to the energy, the total spin
of a state is usually known, the expectation value of the total spin,
$\langle {\bf S}^2 \rangle$, can be used as a measure of how converged
the perturbation expansion is. It can also be used as an indication of
when the Hartree--Fock description is too far away from the physical
situation to be a good enough starting point. This is discussed
further in Sections~\ref{sec:validity} and \ref{sec:result}.

\subsubsection{second--order correction to a Hartree--Fock starting point}

The leading energy correction to the  Hartree--Fock starting point is
of second order in the perturbation (defined in Eq.~(\ref{eq:perturb})).
When $\hat{h}=\hat{h}_s+\hat{u}_{HF}$ and
$\hat{U}=\sum_{i<j}\frac{1}{\hat{r}_{ij}}-\sum_{i=1}^N\hat{u}_{HF}(i)$,
the corresponding corrections to the
wave function will not include any single excitations. This is usually
referred to as Brillouin's theorem and is discussed in standard
Many--Body theory textbooks, see  e.g. Lindgren and Morrison\cite{mbpt}.
Starting from the HF--Hamiltonian for $N$ electrons in the dot we write the
second--order
correction to the energy
\begin{equation}
\delta E_N^{(2)} = \sum_{a<b\leq N}\sum_{\substack{r,s>N \\ r\neq s}}
\frac{|\langle rs|\frac{1}{\hat{r}_{12}}|ab \rangle |^2-
\langle ba|\frac{1}{\hat{r}_{12}}|rs\rangle \langle rs|\frac{1}{\hat{r}_{12}}|ab\rangle}
{\epsilon_a+\epsilon_b-\epsilon_r-\epsilon_s}
\label{perturbation_expansion_eq}
\end{equation}
where thus {\em both} $r$ and $s$ are unoccupied states.

 Since B--splines are used for the expansion of the radial part of the
wave functions there is a finite number of radial quantum numbers ($n$)
to sum over in the second sum of Eq. (\ref{perturbation_expansion_eq}). However, in principle there is still an
infinite number of angular quantum numbers ($m_{\ell}$) to sum over in the same
sum. In praxis this summation has to be truncated and the effects of this
truncation will be discussed in section \ref{sec:validity}.

\subsubsection{Other starting points than Hartree-Fock\label{other starts}}

In principle any starting point, with wave functions close enough to
the true wave functions (to ensure convergence of the perturbation
expansion), can work as a starting point for MBPT. We have in addition
to HF investigated three alternative starting points. If there are
important cancellations between the full exchange (included in
Hartree-Fock) and correlation (not included in Hartree-Fock) an
alternative starting point might converge faster, or even provide
convergence in regions where it cannot be achieved with Hartree-Fock.
First of all we start with the simplest possible starting point; the
pure one--electron wave functions. In this case the basis set consists
of the solutions to the pure 2D harmonic oscillator in the chosen box
and we treat the whole electron--electron interaction as the
perturbation. The second alternative starting point is the Local
Density Approximation (LDA). That is we switch the second term in
Eq.~(\ref{HFoperator}) to $\alpha \langle B_j| 4 a_B^* \sqrt{\frac{2
\rho(r)}{\pi}} | B_i\rangle$, where $\rho(r)$ is the radial electron
density and $\alpha$ is called Slaters exchange parameter and is
usually set to one. Both these starting Hamiltonians are defined with
only local potentials and will thus commute with the total spin. The
third alternative starting point is a reduced exchange HF, i.e. the
exchange term (the second term) in Eq. (\ref{HFoperator}) is simply
multiplied with a constant $\alpha < 1$. When using these alternative
starting points, one must in contrast to the Hartree-Fock case include
single excitations in the perturbation expansion.

The second--order perturbation correction then becomes
\begin{eqnarray}
\delta E_N^{(2)} = \sum_{a,b<N}\sum_{r>N} \frac{|
\langle r| \hat{V}_{ex} |a\rangle - \langle r b|\frac{1}{r_{12}} |b a \rangle
|^2}{\epsilon_a-\epsilon_r}+ \nonumber \\
\sum_{a<b\leq N}\sum_{\substack{r,s>N \\ r\neq s}}
\frac{|\langle rs|\frac{1}{\hat{r}_{12}}|ab \rangle |^2-
\langle ba|\frac{1}{\hat{r}_{12}}|rs\rangle \langle rs|\frac{1}{\hat{r}_{12}}|ab\rangle}
{\epsilon_a+\epsilon_b-\epsilon_r-\epsilon_s}
\end{eqnarray}
where $\hat{V}_{ex}$ is the chosen exchange operator. From this
expression it is also clear that the first term yields zero in the
pure Hartree--Fock case, i.e. then  all single
excitations cancel.

\subsubsection{Full CI treatment of the two body problem}

To investigate how well second--order many-body perturbation theory
performs we have for the simple case of two electrons
also solved the full CI problem. We then diagonalize the matrix that
consists of all the elements of the form
\begin{equation}
H_{ji} = \left\langle
mn\right|_j\:\hat{h}_s^1+\hat{h}_s^2+\frac{1}{\hat{r}_{12}}\:|op\rangle_i
\end{equation}
for given values of $M_L=\sum m_{\ell}$ and $M_S=\sum m_s$ of our electron
pairs $\{|mn\rangle_i\}$. Following the selection rules produced by
Eq. (\ref{ee_interaction_BIG_eq}) we get the conditions
$m_{\ell}^o+m_{\ell}^p=m_{\ell}^m+m_{\ell}^n$, $m_s^m=m_s^o$ and $m_s^n=m_s^p$.

\section{Validation of the Method\label{sec:validity}}

The main purpose of this work is to investigate the usability of many
body perturbation theory on (GaAs) quantum dots. Therfore we have in
this section compared our results with results from other theoretical
works.

Our energies are generally given in meV. For easy comparison with
other calculations it should be noted that the scaled atomic unit for
energy is $1 \, \, \mathrm{Hartree}^* = 1 \, \, \mathrm{Hartree} \,
\left(m^*/(m_e \varepsilon_r^2)\right) \approx 11.857$ meV, with $m^*
= 0.067 m_e$ and $\varepsilon_r=12.4$. The scaled Bohr radius is $
a_B^* = \left(\varepsilon_r m_e/m^* \right) a_B \approx 9.794$ nm.

\subsection{The two electron dot}

\begin{figure}
\includegraphics[width=\figurewidth]{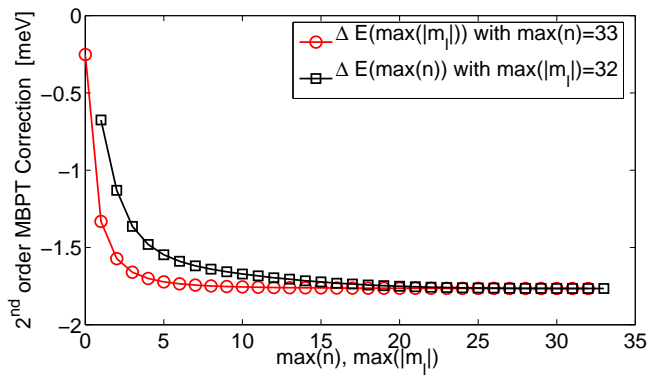}
\caption[]{Second--order perturbation theory correction to the energy
as function of max$(n)$ (squares) and max$(|m_{\ell}|)$ (circles) in
the second sum of Eq.~(\ref{perturbation_expansion_eq}) for the two
electron dot with the confinement strength $\hbar\omega=6$ meV. Note
that the sum over $m_{\ell}$ converges faster than the sum over $n$.
  \label{twoelcompare_numerics}}
\end{figure}

Fig.~\ref{twoelcompare_numerics} shows the second--order many--body
perturbation correction to the energy, Eq.
(\ref{perturbation_expansion_eq}), as function of max$(n)$ (squares)
and max$(|m_{\ell}|)$ (circles) respectively for the two electron dot
with $\hbar\omega=6$ meV. It clearly illustrates that both curves
converge but also that the sum over $m_{\ell}$ converges faster than
the sum over $n$. Due to this we have throughout our calculations used
all radial basis functions and as many angular basis functions that
are needed for convergence. One should, however, notice that the
relative convergence as a function of max$(n)$ and max$(|m_{\ell}|)$
varies with the confinement strength and occupation number. Weak
potentials ($\hbar\omega<2$ meV) usually produce the opposite picture
i.e. a faster convergence for $n$ than for $m_{\ell}$. For confinement
strengths ($\hbar\omega>3$ meV) and most occupation numbers the trend
shown in Fig.~(\ref{twoelcompare_numerics}) is, however, typical.

\subsubsection{Comparison between different starting points}

\begin{figure}
\includegraphics[width=8.6cm]{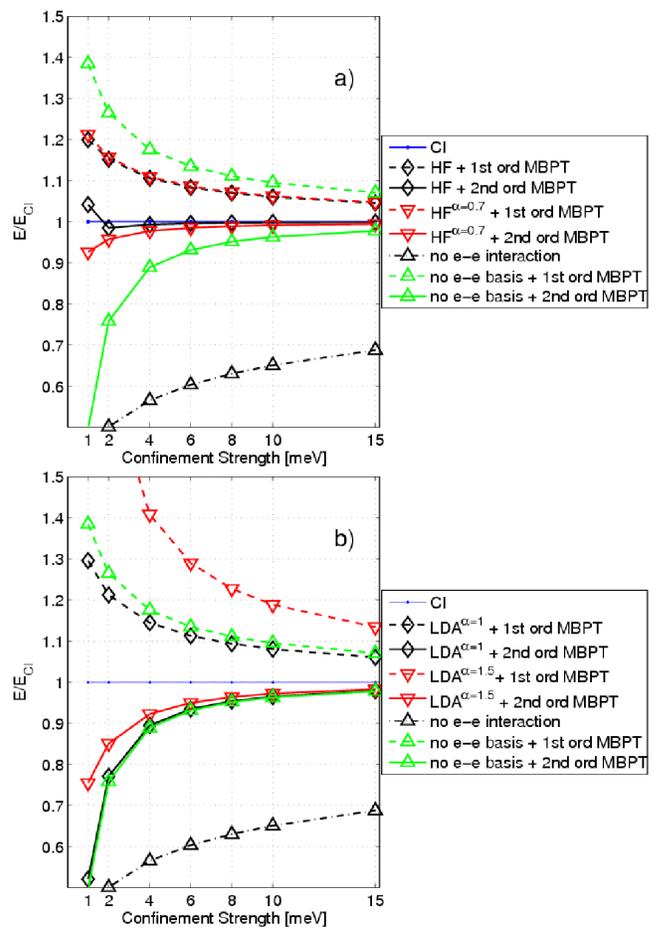}
\caption[]{The quotient between the calculated energies (of respective
  method) and the CI--energy as functions of the confinement strength,
  $\hbar \omega$, for the ground state in the two electron dot. In a)
  the results from HF, a reduced exchange HF with $\alpha=0.7$ and 2nd
  order MBPT using wave functions from respective method are plotted.
  In b) the results from LDA-calculations (with two different alphas) +
  1st and 2nd order MBPT are plotted. For reference the results from
  calculations where we have used the one-electron wave functions as a
  starting point for the perturbation expansion (taking the whole
  electron-electron interaction as the perturbation) have been plotted
  in both a) and b).
\label{twoelcompare}}
\end{figure}

\begin{figure}
  \includegraphics[width=\figurewidth]{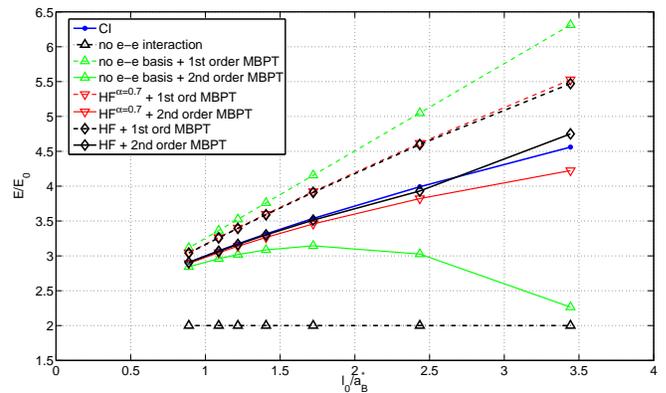}
  \caption[]{ $E/E_0$ for the two-electron dot and for different
methods as functions of $l_0/a_B^*$, where $l_0=\sqrt{\hbar / (m^*
\omega )}$ is the characteristic dot length, $E_0 = \hbar \omega$ is the
single particle energy, and $a_B^*$ is the effective Bohr radius.
Small values of $l_0/a_B^*$ correspond to stronger confinement and therefore
a faster expected convergence rate of a perturbation expansion.
    \label{twoelcompareDIMLESS}}
\end{figure}

In Fig.~\ref{twoelcompare} a) and b) comparisons between HF (i.e. the
expectation value of the full Hamiltonian with a Slater determinant of
Hartree-Fock orbitals, labelled HF + 1st order MBPT in Figs.
\ref{twoelcompare} and \ref{twoelcompareDIMLESS}), the alternative starting
points discussed in section~\ref{other starts}, second--order MBPT (HF
or alternative starting point + second--order correlation) and CI
calculations for the ground state in the two electron dot are made.
Both the second--order MBPT and CI results have been produced with all
radial basis functions ($33$ for each combination of $m_{\ell}$ and
$m_s$) and $-6\le m_{\ell}\le 6$. It is clear from
Fig.~\ref{twoelcompare}a) that second--order correlation here is the
dominating correction to the Hartree-Fock result. Even for
$\hbar\omega = 2$ meV the difference compared to the CI result
decreases with one order of magnitude when it is included. For
stronger confinements the difference to CI is hardly visible. As
expected, the performance of both HF and second--order MBPT is
improved when stronger confinement strengths are considered. For the
weakest confinement strength calculated here ($\hbar\omega = 1$ meV)
the pure Hartree--Fock approximation gives unphysical wave functions
in the sense that the spin up and the spin down wave functions differ,
resulting in a non--zero $\langle S^2 \rangle$. This shows up in
figure a) as a broken trend (all of a sudden an overestimation of the
energy instead of an underestimation) for the pure HF + second--order
correlation curve at $\hbar\omega=1$ meV. For all other potential
strengths $\langle S^2 \rangle$ is zero to well below the numerical
precision ($ \backsim 10^{-6}$) for both the Hartree-Fock and
second--order MBPT wave functions, while for the $\hbar\omega=1$ meV
calculation $\langle S^2 \rangle=0.33$ and $0.26$ for the Hartree-Fock
and second--order MBPT calculations respectively. It should be noted
that at $\hbar\omega=1$ meV the energy of the second--order MBPT
calculation is still only 4$\%$ larger than the CI-value (although the
wave functions are unphysical) and that probably the state will
converge to $\langle S^2 \rangle =0$ when MBPT is performed to all
orders. All other tested starting points yield $\langle S^2 \rangle
=0$ for this confinement strength, but still their energy estimates
after second--order MBPT are worse. This shows that conserved spin
does not necessary yield good energies and broken spin symmetry does
not necessary yield bad energy estimations. We note that the reduced
exchange Hartree--Fock, displayed in Fig.~\ref{twoelcompare}a) , seems
to be a fruitful starting point for perturbation theory although the
results after second--order are slightly worse than after the full
exchange Hartree--Fock + second--order MBPT. For $\hbar\omega=1$ meV
the reduced exchange HF with $\alpha=0.7$ still gives $\langle S^2
\rangle=0$, i.e. the onset of spin contamination is delayed on the
expense of proximity to the CI-energy. To put it in another way, if we
lower $\alpha$, the corresponding curve in Fig \ref{twoelcompare} a)
will be lower (and thus further from the correct CI--curve), but the
spin contamination onset will appear for a weaker confinement
strength. This freedom could be useful when doing MBPT to all orders.

From Fig. \ref{twoelcompare} b) we conclude that LDA with $\alpha > 1$
might be a useful starting point for perturbation theory calculations
to all orders but not a good option for 2nd order calculations, at
least not for weak potentials. LDA calculations with $\alpha =1$,
however, seems to be a bad starting point for MBPT, at least for the
two electron case, since it gives almost identical results after
second order as using the pure one electron wave functions as starting
point. LDA might still work better as a starting point when more
electrons are added to the dot.

Finally the comparison with the pure one--electron wave functions in
Fig.~\ref{twoelcompare} clearly illustrates how much of an improvement
it is to start the perturbation expansion from wave functions that
already include some of the electron--electron interaction, especially
for weaker potentials. This becomes even more clear in
Fig.~\ref{twoelcompareDIMLESS} where we present the results from
Fig.~\ref{twoelcompare} a) in another way. Here we have plotted
$E_{Method}/E_{0}$ as functions of $l_0/a_B^*$ where $E_0=\hbar\omega$
is the single particle energy and $l_0=\sqrt{\hbar/(m^*\omega)}$ is
the characteristic length of the dot. It demonstrates what an
extraordinary improvement it is to start from Hartree--Fock compared
to starting with the one--electron wave functions when doing
second--order MBPT for low electron densities (high $l_0/a_B^*$). It
also seems as there is a region where the Hartree-Fock starting point
would yield a convergent perturbation expansion while taking the whole
electron--electron interaction as the perturbation would not.

\subsection{The six electron dot}

\begin{figure}[h!]
\includegraphics[width=\figurewidth]{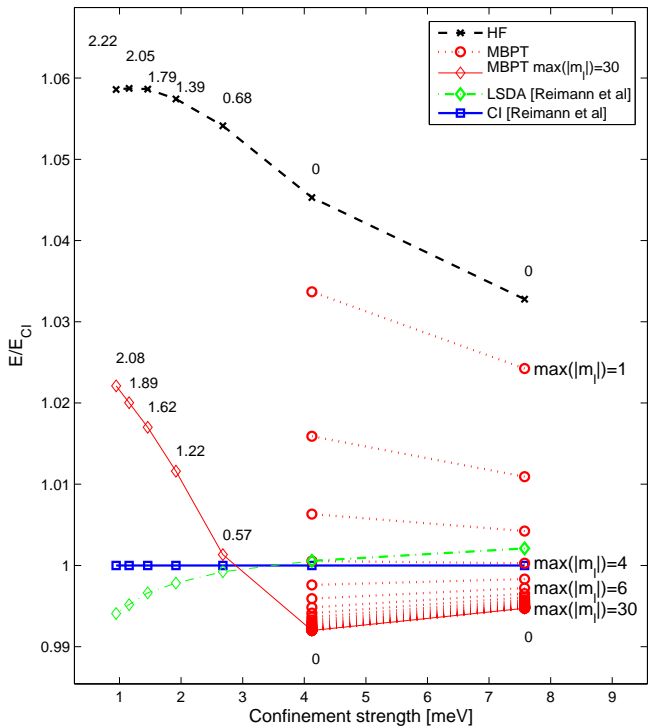}
\caption[]{Comparison between our HF and second--order MBPT results
for the six electron dot in the ground state with $M_L^{TOT}=0$ and
$S_z^{TOT}=0$, with the LSDA and CI calculations by Reimann {\it et
al}\cite{ReimannCI}. The second--order MBPT calculations include the
full sum over the complete radial basis set (corresponding to all
$n$-values) and with max$(|m_{\ell}|)=1,2,3,\ldots,30$ for the two
strongest potentials. For clarity only the curves with
max$(|m_{\ell}|)=1,4,6$ and $30$ have been labeled. The HF and the
second--order MBPT with max$(|m_{\ell}|)=30$ curves are plotted for
all potential strengths calculated by Reimann et al. Moreover, the
values of $\langle S^2 \rangle$ for the HF and the second--order MBPT
with max$(|m_{\ell}|)=30$ have been plotted in the
figure.\label{sixelcompare}}
\end{figure}

In Fig.~\ref{sixelcompare}, a comparison between our HF and
second--order MBPT calculations on the ground state of the six
electron dot is made with a DFT calculation in the Local Spin Density
Approximation(LSDA) as well as with a CI calculation, both by Reimann
{\it et al.}~\cite{ReimannCI}. They performed their
calculations for seven different electron densities here translated
to potential strengths. Let us first focus on the results for the two
highest densities, corresponding to a Wigner-Seitz radius $r_s = 1.0
a_B^*$ and $r_s = 1.5 a_B^*$ which translates to confinement strengths
of $\hbar \omega \approx 7.58$ meV and $\approx 4.12$ meV
respectively. The reason that we want to separate the comparison for
those confinement strengths is that our Hartree-Fock calculations
yield solutions with $\langle S^2 \rangle >0$ for the weaker
confinement strengths. A similar behavior was seen by Sloggett {\em et
al.}~\cite{Slogget} in their unrestricted HF calculations. Therefore
the results for the weaker potentials overestimate the energy in a
unphysical manner;  compare the above discussion around Fig.
\ref{twoelcompare} a). The CI-method however always yield $\langle S^2
\rangle = 0$ for the closed 6 electron shell and consequently a
comparison with spin contaminated results would here, in some sense,
be misleading. It should be emphasized that the spin contamination is
a feature of our choice of starting point and not a problem with MBPT
in itself.

 To make comparison easy all energies are normalized to the
corresponding CI--value. The figure clearly illustrates, for the two
stronger confinement strengths, that while the HF results overshoot
the CI energy by between $3.5\%$ and $4.5\%$ the second order MBPT
calculations improve the results significantly. Already for
max$(|m_{\ell}|)=1$ the energy only overshoots the CI value with between
$2.5\%$ and $3.5\%$ while the second--order MBPT energy for
max$(m_{\ell})=4$ is almost spot on the CI energy. However, with
max$(m_{\ell})=30$ the second--order MBPT gives somewhere between $0.5\%$
and $1\%$ lower energy than the CI calculation. We note that the CI
calculation by Reimann {\it et al.} was made with a truncated basis set
consisting of the states occupying the eight lowest harmonic
oscillator shells. This means e.g. that their basis set includes only
two states with $(|m_{\ell}|)=5$ and one with $(|m_{\ell}|)=6$. Within this
space all possible six electron determinants were formed. After
neglecting some determinants with a total energy larger than a chosen
cutoff, the Hamiltonian matrix was constructed and diagonalized.
Fig.~\ref{sixelcompare} indicates that the basis set used in
Ref.~\cite{ReimannCI} was not saturated to the extent probed here,
since almost all interactions with $|m_{\ell}| > 4$ were neglected. According
to Reimann {\it et al.} they used a maximum of $108$ $375$ Slater
determinants while we, through perturbation theory, use a maximum of
$980$ $366$ Slater determinants. The difference of our max$(|m_{\ell}|)=30$
results and their CI results are thus not unreasonable. Since Reimann
{\it et al.} solved the full CI problem, the matrix to diagonalize is
huge and it is, according to the authors, not feasible to use an even
larger basis set. An alternative could be to include more basis
functions, but restrict the excitations to single, doubles and perhaps
triples. The domination of double excitations is well established in
atomic calculations, see e.g. the discussion in Ref.~\cite{amp:91:be}.
It should however be noted that the difference between the results
concerns the fine details. Our converged results are less than one
percent lower than those of Reimann {\it et al.} and when using
approximately the same basis set as they did (max$(|m_{\ell}|)=4$) the
difference between the results is virtually zero. Moreover, we see for
the two strongest potentials the same trend as we saw in the
two--electron case, namely that the HF, MBPT and CI results tend
towards one another with increasing potential strength. This trend is
not seen for the LSDA approach.

Finally, Fig.~\ref{sixelcompare} shows, for the five weakest potentials,
that our HF results get increasingly spin contaminated when the
potential is weakened. Hereby the HF--approximation artificially lowers
its energy and subsequently this leads to an overestimation of the
second--order MBPT energies for these potential strengths.
Surprisingly, however, the energy is never more than just above $2\%$
over the CI--results even when $\langle S^2 \rangle>2$. Note also that
MBPT improves the HF--value of $\langle S^2 \rangle$ as it should.

\subsection{Correlation in an external Magnetic Field}

The behavior of quantum dots in an external magnetic field applied
perpendicular to the dot has previously been examined many times both
experimentally e.g.\cite{Tarucha,Kouwenhoven,TaruchaMagenticWeakPot}
and theoretically e.g~\cite{Tarucha,SteffensMagneticCSDFT,Szafran}. The
chemical potentials $\mu(N)=E(N)-E(N-1)$ plotted versus the magnetic
field usually show a rich structure, including e.g. state switching and
occupation of the lowest Landau band at high magnetic fields.

\begin{figure}[h]
\includegraphics[width=\figurewidth]{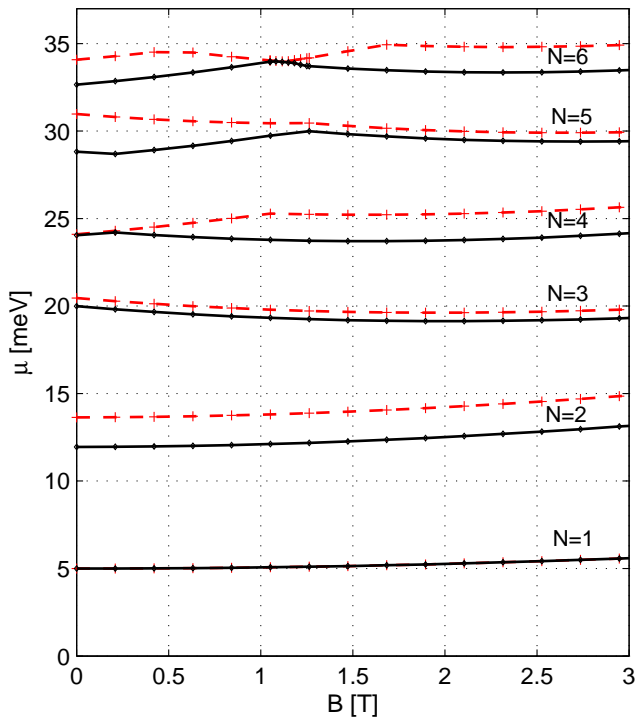}
\caption[]{The chemical potentials for $N=1,2,$\ldots$,6$ as functions
of the external magnetic field according to HF (dashed curve) and
second--order MBPT (full curve) calculations for the potential
strength $\hbar\omega=5$ meV. Note the big difference between the two
different models regarding the behavior of the chemical potentials when the
magnetic field varies.\label{Bfield}}
\end{figure}

Fig.\ref{Bfield} shows the chemical potentials for $N=1,2,\ldots,6$ as
functions of the magnetic field according to our HF (dashed curves)
and second--order MBPT with $-10\le m_{\ell}\le10$ (full curves)
calculations for the potential strength $\hbar\omega=5$ meV. We have
here limited ourselves to the first six chemical potentials calculated
at selected magnetic field strengths (shown by the marks in the
figure). We emphasize again that our intention here is rather to test
the capability of MBPT in the field of quantum dots than to provide a
true description of the whole experimental situation. With increasing
particle number MBPT naturally becomes more cumbersome, but magnetic
field calculations are feasible at least up to $N=20$.

First note the significant difference between the HF and 
second--order MBPT results. Once again correlation proves to be extremely
important in circular quantum dots. With our correlated results we
also note a close resemblance both to the experimental work by Tarucha
{\em et al.}~\cite{Tarucha} and to the current spin-density calculation by Steffens {\em et
al.}~\cite{SteffensMagneticCSDFT}, made with the same potential and material parameters as
used here. (Note that Ref.~\cite{SteffensMagneticCSDFT} defines the
chemical potentials as $\mu(N)=E(N+1)-E(N)$, shifting all curves one
unit in $N$). An example of the importance of correlation is the
four-electron dot that switches state from $|\sum_{i=1}^N n_i,|M_L|,S\rangle=|0,0,1\rangle$ to
$|0,2,0\rangle$ at approximately $1$ T in the HF calculations and at
approximately $0.2$ T in the correlated calculations. We want to
emphasize that we have found the exact position of this switch to be
very sensitive to the potential strength and to the value of $g^*$.
The big difference concerning the magnetic field where this switch occurs
can probably be attributed to the HF tendency to strongly favor
spin-alignment. This is an effect originating from the inclusion of
full exchange, but no correlation. Inclusion of second--order
correlation energy cures this problem. Finally we note that the $N=5$
switch from $|0,1,\frac{1}{2}\rangle$ to $|0,4,\frac{1}{2}\rangle$ in
our correlated calculations takes place somewhere around $1.2$ T which
is also in agreement with both mentioned studies.

\section{Results\label{sec:result}}

\subsection{The addition energy spectra}

\begin{figure}
{
\includegraphics[width=\figurewidth]{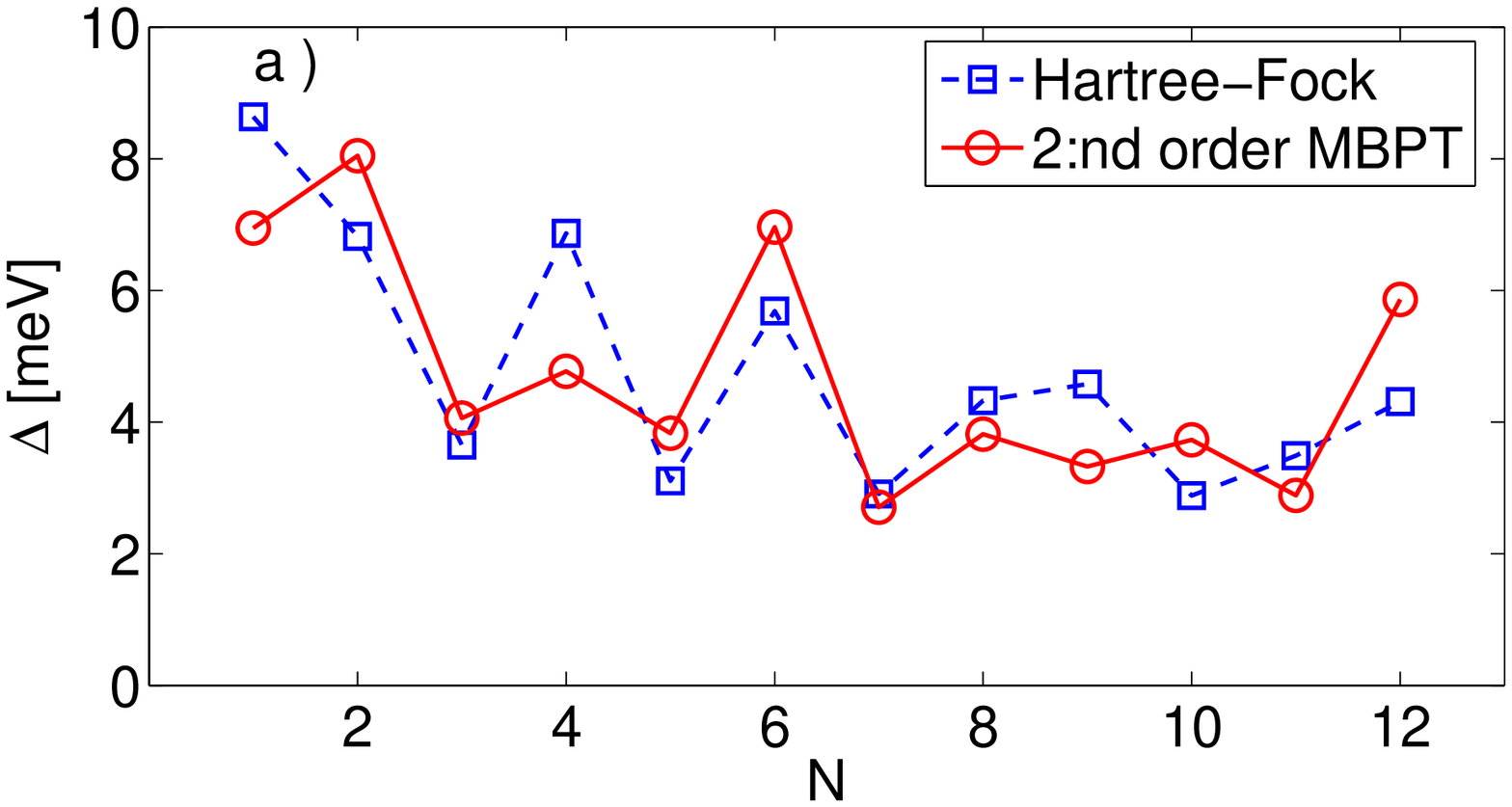}}
{
\includegraphics[width=\figurewidth]{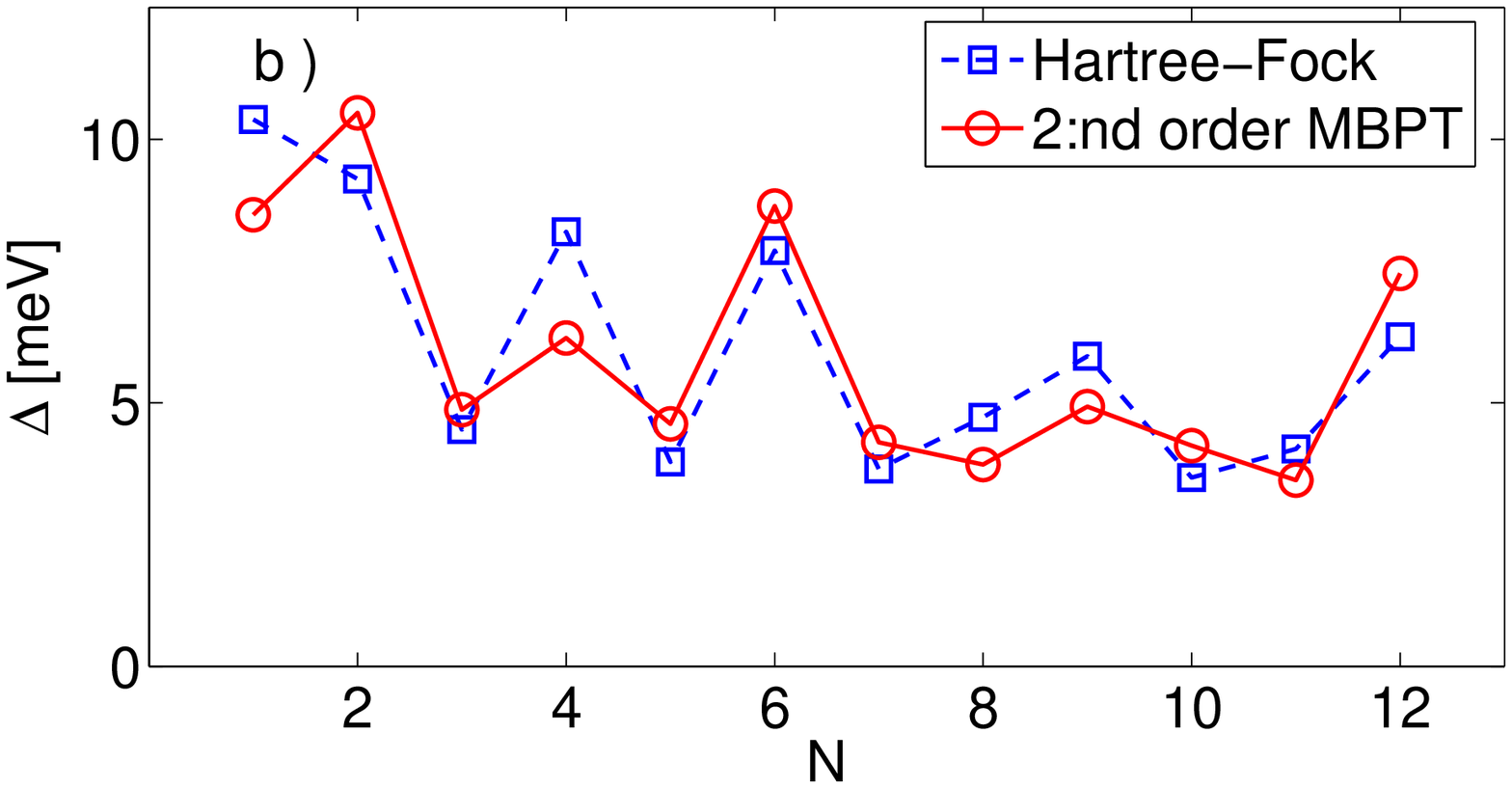}
} \caption[]{\label{adspec}The ground state addition energy spectra for dots with
  $\hbar\omega=5$ meV (a) and $\hbar\omega=7$ meV (b). The squares
  (circles) represent the addition energy spectra according to HF
  (second--order MBPT). It is clear that the second--order
  MBPT--spectra imply closer resemblances to the experimental
  picture in Tarucha\cite{Tarucha} than the HF--spectrum.
  }
\end{figure}

The so called addition energy spectra, with the addition energy
defined as $\Delta(N)=E(N+1)-2E(N)+E(N-1)$, have been widely used to
illustrate the shell structure in quantum dots. Main peaks at
$N=2,6,12$ and $20$, indicating closed shells, and subpeaks at $N=4,9$
and $16$, due to maximized spin at half filled shells, have been
interpreted as the signature for truly circular quantum
dots~\cite{Reimanncomparetarucha}. Experimental deviations from this
behavior have been interpreted as being due to nonparabolicities of the
confining potential or due to 3D--effects\cite{MatagneRealisticVQD}.
We here show that correlation effects in a true 2D harmonic potential
can in fact generate an addition energy spectrum with similar
deviations.

In this work we limit ourselves to the first three shells since it
seems as the experimental situation is such that the validity of the
2D harmonic oscillator model becomes questionable with increasing
particle number\cite{MatagneRealisticVQD}. Calculations of dots with
larger $N$ could, however, readily be made with our procedure. The
addition energy spectra are produced with $-10\le m_{\ell}\le 10$. The
filling order for the first six electrons is straight forward. When
the seventh electron is added to the dot the third shell starts to
fill. With a pure circular harmonic oscillator potential and no
electron-electron interaction the $\mid 0, \pm 2, \pm \frac{1}{2}
\rangle$ and $\mid 1, 0, \pm \frac{1}{2} \rangle$ one-particle states
are completely degenerate. This degeneracy is lifted by the
electron-electron interaction, but not more than that the energies
have to be studied in detail in order to determine the filling order.
Similar conclusions, that the filling order is very sensitive to small
perturbations, have been drawn by Matagne {\em et
al.}\cite{MatagneRealisticVQD}, who studied the influence of
non-harmonic 3D effects. Our focus is instead the detailed description
of the electron-electron interaction. For $N = 7-11$ we have thus
calculated all third shell configurations, and for each configuration
considered the maximum spin. The results are found in
Table~\ref{tab:energies}. For each number of electrons we can identify
a ground state, which sometimes differs between HF and MBPT. These
ground states are used when creating Fig.~\ref{adspec}a) and b). The
energy gap to the first excited state is sometimes very small and the
possibility of alternative filling orders will be discussed in the
next section.

Fig.~\ref{adspec}a) and Fig.~\ref{adspec}b) thus show the ground state
addition energy spectra up to $N = 12$ according to the Hartree-Fock
model as well as to second--order MBPT for $\hbar\omega=5$ and $7$ meV. 
Note first the big difference between the HF and MBPT spectra. 
These figures clearly illustrate how
important correlation effects are in these systems. Admittedly the
HF--spectra show peaks at $N=4,6,9$ and $12$ but the relative size of
the addition energy between closed and half--filled shells is not
consistent with the experimental
picture\cite{Tarucha,MatagneRealisticVQD}. The second--order
MBPT--spectra have in contrast clear main peaks at $N=2,6$ and $12$,
indicating closed shells, and a $N=4$ subpeak indicating maximized
spin for the half filled shell. For the $\hbar\omega=7$ meV spectrum
the subpeak at $N=9$ is also clear but for the $\hbar\omega=5$ meV
spectrum the subpeak at $N=9$ is substituted by subpeaks at $N=8$ and
$10$. The behavior of the addition energy spectra in this, the third
shell, will be discussed in detail below.

\begin{table*}[ht!]
\caption{\label{tab:energies} Energy of the ground and third shell
excited state for 7--11 electron dots with $\hbar\omega=5$ and $7$ meV.
The notation $(\sum_{i=1}^N n,|M_L|,S)$ to label the state is used.
The ground state energy according to Hartree--Fock (HF energy) and to HF +
second--order MBPT (Correlated energy) and for
respective N and potential strength is marked in bold.}
\begin{ruledtabular}
\begin{tabular}{l|l|cccc|cccc}
\hline
\# $e^{-}$ & & \multicolumn{4}{c|}{$\hbar\omega=5$ meV} & 
\multicolumn{4}{c}{$\hbar\omega=7$ meV}\\
\hline
7 & State  & \multicolumn{2}{c}{$(0,2,\frac{1}{2})$} & 
\multicolumn{2}{c|}{$(1,0,\frac{1}{2})$} & 
\multicolumn{2}{c}{$(0,2,\frac{1}{2})$} & 
\multicolumn{2}{c}{$(1,0,\frac{1}{2})$} \\
\hline
 & HF energy [meV]  &\multicolumn{2}{c}{\bf{168.02}}&
\multicolumn{2}{c|}{168.67} &\multicolumn{2}{c}{$\mathbf{215.80}$} & 
\multicolumn{2}{c}{$216.58$}\\
 & Correlated energy [meV] &\multicolumn{2}{c}{\bf{162.08}}&
\multicolumn{2}{c|}{162.15} &\multicolumn{2}{c}{$\mathbf{208.96}$} & 
\multicolumn{2}{c}{ $209.52$}\\
 \hline
8&State & $(0,0,1)$& $(0,4,0)$&$(1,2,1)$& $(2,0,0)$& $(0,0,1)$& $(0,4,0)$&$(1,2,1)$& $(2,0,0)$\\
\hline
&HF energy [meV] & \bf{210.69}&212.33 &211.66 &214.00 & $\mathbf{270.66}$ & 272.20&  271.51& 274.32\\
&Correlated energy [meV] &205.23 &\bf{204.40} &204.66 &205.02 &263.82 &$\mathbf{263.70}$& 263.85& 264.65\\
\hline
9&State & $(1,0,\frac{3}{2})$&$(0,2,\frac{1}{2})$ &
$(1,4,\frac{1}{2})$&$(2,2,\frac{1}{2})$&
$(1,0,\frac{3}{2})$&$(0,2,\frac{1}{2})$ & $(1,4,\frac{1}{2})$ &$(2,2,\frac{1}{2})$ \\
\hline
&HF energy [meV]& \bf{257.69}&259.24 &259.28 & 260.56&$\mathbf{330.25}$ & 332.17&  332.14& 333.64\\
&Correlated energy [meV] &\bf{250.54} &251.35 & 250.95&251.00 & $\mathbf{322.27}$ & 322.81&  323.06& 323.37\\
\hline
10&State & $(1,2,1)$& $(0,0,0)$& $(2,0,1)$& $(2,4,0)$& $(1,2,1)$& $(0,0,0)$&$(2,0,1)$& $(2,4,0)$\\
\hline
&HF energy [meV]&\bf{309.27} & 310.64& 310.17&311.06&$\mathbf{395.72}$ & 397.20&396.73& 397.78 \\
&Correlated energy [meV] &300.49 &\bf{300.00} & 300.25&300.52& 385.92& $\mathbf{385.76}$& 386.06& 386.49\\
\hline
11&State & \multicolumn{2}{c}{$(1,0,\frac{1}{2})$}&
\multicolumn{2}{c|}{$(2,2,\frac{1}{2})$}& 
\multicolumn{2}{c}{$(1,0,\frac{1}{2})$}& \multicolumn{2}{c}{$(2,2,\frac{1}{2})$}\\
\hline
&HF energy [meV]&\multicolumn{2}{c}{\bf{363.72}} & 
\multicolumn{2}{c|}{364.49} &\multicolumn{2}{c}{$\mathbf{464.77}$}& 
\multicolumn{2}{c}{465.57}\\
&Correlated energy [meV] &\multicolumn{2}{c}{353.66} &
\multicolumn{2}{c|}{\bf{353.19}} &\multicolumn{2}{c}{453.47}& 
\multicolumn{2}{c}{$\mathbf{453.43}$}\\
\hline
\end{tabular}
\end{ruledtabular}
\end{table*}

\begin{table*}
\caption{Expectation values of $S^2$ for the cases where correlation
switches ground states in the third shell. The state labeled ``Ground
State'' is the ground state according to second--order MBPT while the
state labeled ``Excited State'' is the ground state according to
Hartree-Fock but an excited state according to second--order MBPT.
\label{tab:Spin} }
\begin{ruledtabular}
\begin{tabular}{ll|cl|cl|cl|cl}
\hline 
\# e$^{-}$ &         &\multicolumn{4}{c|}{$\hbar\omega=5
  meV$}&\multicolumn{4}{c}{$\hbar\omega=7 meV$}\\
\hline
 & &\multicolumn{2}{c|}{Ground State}&\multicolumn{2}{c|}{Excited state} &\multicolumn{2}{c|}{Ground State}&\multicolumn{2}{c}{Excited state}\\
\hline
\multicolumn{2}{l|}{~} & E [meV] & $\langle S^2 \rangle $ & E [meV] & $\langle S^2
\rangle $ & E [meV] & $\langle S^2 \rangle $ & E [meV]& $\langle S^2 \rangle $\\
\hline
$8$  & HF                &212.33&0.00&210.69&2.70&272.20&0.00&270.66&2.30 \\
     &$2^{nd}$--ord MBPT &204.40&0.00&205.23&2.58&263.70&0.00&263.82&2.22\\
     &Exact              &      &   0&      &2   &      &0   &      &2\\       
 \hline
$10$  & HF                         &310.64&0.00&309.27&2.21& 397.20& 0.00 &395.72&2.08 \\
     &$2^{nd}$--ord MBPT &300.00&0.00&300.49&2.15&385.76 & 0.00 &385.92&2.05 \\
     &Exact              &      &   0&      &2   & &0 &      &2    \\       
 \hline
$11$  & HF               &364.49&0.77&363.72&0.99&465.57&0.758&464.77&0.82\\
      &$2^{nd}$--ord MBPT&353.19&0.76&353.66&0.93&453.43&0.755&453.47&0.79\\
      &Exact             &      &0.75&      &0.75&      &0.75 &      &0.75 \\
 \hline
\end{tabular}
\end{ruledtabular}
\end{table*}

\subsubsection{Filling of the third shell}

The filling of the third shell has previously been examined by Matagne
{\em et al.}~\cite{MatagneRealisticVQD} both experimentally and
theoretically. In their theoretical description they use a 3D DFT model
with the possibility to introduce a nonharmonic perturbation that can
change the ground states in the third shell and thereby alter the
addition energy spectra. They then compare their theoretical
description with different experimental addition energy spectra and
argue how large deviation from the circular shape they have in the
different experimental setups. They conclude that a clear dip at $N=7$
followed by a peak at $N=8$ or $9$ is a signature of maximized spin at
half filled shell and that a dip at $N=7$ and the filling sequence
\begin{eqnarray}
\label{MatagneFillingorder}
\left| \sum_{i=7}^N n^i, |\sum_{i=7}^N m_{\ell}^i|, S \right\rangle =
\, \mid 0, 2, \frac{1}{2} \rangle \Rightarrow
\mid 0, 0, 1 \rangle \Rightarrow \nonumber \\
\mid 1, 0, \frac{3}{2} \rangle \Rightarrow
\mid 1, 2, 1 \rangle \Rightarrow
\mid 1, 0, \frac{1}{2} \rangle \Rightarrow
\mid 2, 0, 0 \rangle
\end{eqnarray}
for the six electrons to enter the third shell is a signature of a
``near ideal artificial atom''. This is also the filling sequence we
find using the HF- approximation. As seen in Fig.~\ref{adspec}a) and
b) there is then indeed also a dip at $N=7$ and a peak at $N=9$. The
dip at $N=7$ is further supported by the DFT calculation by Reimann
{\em et al.}~\cite{Reimanncomparetarucha}. In contrast the experiment
by Tarucha {\em et al.}\cite{Tarucha} did not show the $N=7$ dip. In
Ref.~\cite{MatagneRealisticVQD} this is explained by deviations from
circular symmetry for the specific dot used in Ref.~\cite{Tarucha}. As
will be seen below our many-body calculations give in several cases
different ground states and thus favor a different filling order than
Eq.\ref{MatagneFillingorder}.

Table \ref{tab:energies} shows the ground state and excited
states energies of the third shell according to HF and
second--order MBPT for $\hbar\omega=5$ meV and
$\hbar\omega=7$ meV. Notice that the different methods yield
different ground states for the $8,10$ and $11$--electron systems
although both potential strengths yield the same ground states. Note also
 the small excitation gap between the correlated ground and first
excited state that occurs in some cases. For example between the
$\mid 0,2,\frac{1}{2} \rangle$ and $\mid 1,0,\frac{1}{2} \rangle$
seven-electron states in the
$\hbar\omega=5$ meV dot the energy difference is $0.07$ meV, between
the $\mid 0,0,1 \rangle$ and $\mid 0,4,0 \rangle$
eight-electron state in the $\hbar\omega=7$
meV dot the energy difference is $0.12$ meV and between the
$\mid 1,0,\frac{1}{2} \rangle$ and $\mid 2,2,\frac{1}{2} \rangle$
eleven-electron states in the
$\hbar\omega=7$ meV dot the energy difference is only $0.04$ meV. The
$(1,0, \frac{3}{2})$ state at $N=9$  seems, however, relatively stable for
both potential strengths with excitation gaps of $0.41$ and $0.54$
meV. Surprisingly for both the $\hbar\omega=5$ and $7$ meV the calculations 
including correlations indicate the ground state third shell filling sequence
\begin{eqnarray}
\label{CorrelatedFillingorder}
\mid~0,2,\frac{1}{2} \rangle
\Rightarrow \mid~0,4,0 \rangle
\Rightarrow
\mid~1,0,\frac{3}{2}\rangle \Rightarrow \nonumber \\
\mid~0,0,0 \rangle
\Rightarrow
\mid~2,2,\frac{1}{2} \rangle
\Rightarrow \mid~2,0,0 \rangle
\end{eqnarray}
 for $N = 7-12$. Note that this sequence implies a spin-flip of the
electrons already in the dot when the ninth and tenth electrons are
added. Only the seven-electron dot and the nine-electron dot here have
the same ground state as in HF (whose filling sequence coincides with
that preferred in Ref~\cite{MatagneRealisticVQD}). Matagne {\em et
al.} also discuss that the behavior of the dot examined in
Ref.~\cite{Tarucha} for small magnetic fields implies the sequence
\begin{eqnarray}
\label{TaruchaFillingorder}
\mid~0,2,\frac{1}{2} \rangle \Rightarrow
\mid~0,4,0 \rangle \Rightarrow
\mid~1,2,\frac{1}{2}\rangle \Rightarrow \nonumber\\
 \mid~0,0,0 \rangle \Rightarrow
\mid~1,0,\frac{1}{2} \rangle \Rightarrow
\mid~2,0,0 \rangle,
\end{eqnarray}
 but tend to attribute this to deviations from circular shape. This
filling sequence is indeed much closer to the ground states we have
obtained with a perfect circular potential. This indicates the
possibility that many-body effects usually neglected could have an
effect similar to that of imperfections in the dot construction. We
note in passing that Sloggett and Sushkov~\cite{Slogget} support our
finding of a spin-zero ground-state for ten electrons, although their
calculation was done with a stronger potential. The different
configurations for nine electrons in Eq.~\ref{CorrelatedFillingorder}
and Eq.~\ref{TaruchaFillingorder} can be due to the fact that the experimental
situation favors population of an excited state since population of
the ground state would require a spin flip. However, if we produce a
spectrum with this filling sequence, we get a large dip at $N=9$.
Similarly, when the eleventh electron is injected, the population of our
ground state would require a configuration change of the electrons
already in the dot. 

\begin{figure}
\includegraphics[width=7.5cm]{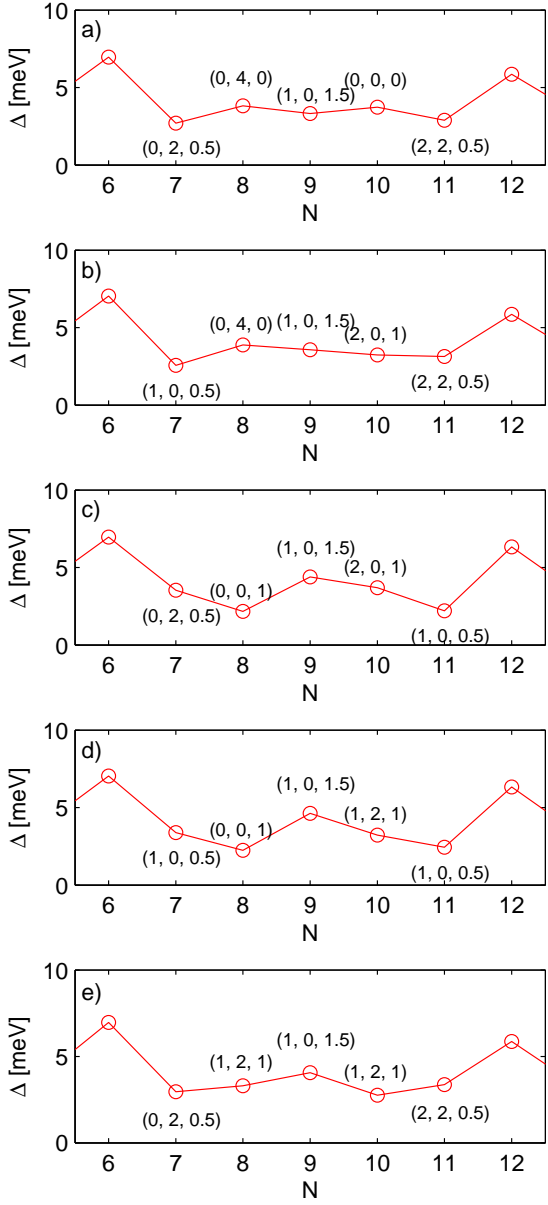}
\caption[]{Ground state, a), and selected excited state, b)-e),
  addition energy spectra for $\hbar\omega=5$ meV according to
  second--order MBPT. The notation $(\sum_{i=7}^N n^i,|\sum_{i=7}^N
  m_{\ell}^i|,S)$ to label the states is used. Note the big differences
  between the different spectra. For example the ground state
  spectrum, a), has peaks at $N=8,10$ while spectrum b) has a peak at
  $N=8$ and the rest have a peak at $N=9$. That is, even if the
  spin is maximized at half filled shell ($N=9$) there is not always a
  peak there as seen in subfigure a) and b). Subfigure e) resembles the
  experimental results of Ref.~\cite{MatagneRealisticVQD} best with
  dips at $N=7$ and $10$ and a peak at $N=9$. Moreover, combining the
  addition energies for $N=6,7,8$ of sequence c) or d) with the
  addition energies for $N=10,11,12$ of sequence e) would give a
  spectrum that closely resembles the experimental situation in
  Ref.~\cite{Tarucha} with dips at $N=8$ and $10$ and a peak at $N=9$.
  \label{AdSpecThirdshell_5}}
\end{figure}

\begin{figure}
\includegraphics[width=7.5cm]{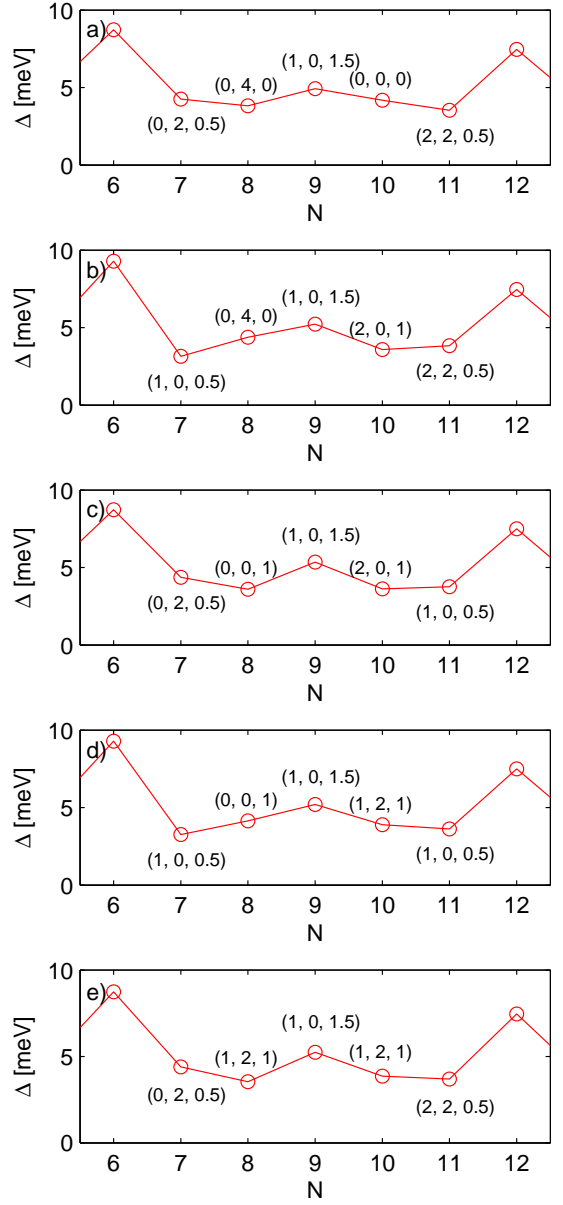}
\caption[]{Ground state, a), and selected excited state, b)-e),
  addition energy spectra for $\hbar\omega=7$ meV according to
  second--order MBPT. The notation $(\sum_{i=7}^N n^i,|\sum_{i=7}^N
  m_{\ell}^i|,S)$ to label the states is used. Note the big differences
  between the different spectra. Note also that all spectra have peaks
  at $N=9$. Even though the ground state spectra for $N=7,8$ and $9$
  resemble the experimental results of Ref.~\cite{Tarucha}, the dip at
  $N=11$ is uncharacteristic when compared with the experimental
  results of Ref.~\cite{Tarucha} and
  Ref.~\cite{MatagneRealisticVQD}. Subfigure b) resembles the
  experimental result in Ref.~\cite{MatagneRealisticVQD} the most with
  a peak at $N=9$ and dips at $N=7$ and $10$ while subfigure c)
  resembles the experimental results of Ref.~\cite{Tarucha} the most
  with a peak at $N=9$ and dips at $N=8$ and $10$.\label{AdSpecThirdshell_7}}
\end{figure}

In Fig.~\ref{AdSpecThirdshell_5} and Fig.~\ref{AdSpecThirdshell_7}
addition energy spectra are shown assuming different filling orders
for $5$ meV and $7$ meV, respectively. In each figure the calculated
ground state filling sequence is shown in the uppermost panel, labeled
a), and then the other panels, e) -- f), show selected excited state
filling sequences. Note that even though the same filling sequences
are used in Fig.~\ref{AdSpecThirdshell_5} and
Fig.~\ref{AdSpecThirdshell_7} the addition energy spectra differ
 between these rather close potential strengths. We can thus
 conclude that a given filling sequence does not yield a unique
addition energy spectra since the relative energies of the ground and
excited states are very sensitive to the exact form of the potential.
Furthermore we agree with Matagne {\em et
al.}~\cite{MatagneRealisticVQD} that full spin alignment for the
nine-electron ground state does not guarantee a peak in the addition
energy spectrum as seen in Fig.~\ref{AdSpecThirdshell_5} a) and b).
Moreover we see that the spectra that resemble the experimental one
in Fig. 3a) of Ref.~\cite{MatagneRealisticVQD} (a clear dip at $N=7$
and $10$ and a clear peak at $N=9$) are Fig.~\ref{AdSpecThirdshell_5}e)
and Fig.~\ref{AdSpecThirdshell_7}b). Finally we see that
Fig.~\ref{AdSpecThirdshell_7}c) resembles the experimental situation
in Ref.~\cite{Tarucha} (dips at $N=8$ and $10$ with a peak at $N=9$)
the most. We certainly do not claim that these filling sequences are
those really obtained in the mentioned experiments. However, we want to
stress that great care must be taken when conclusions are drawn from
comparisons between theoretical and experimental addition energy spectra.

\subsubsection{Spin contamination in the third shell}

\begin{figure}
\includegraphics[width=\figurewidth]{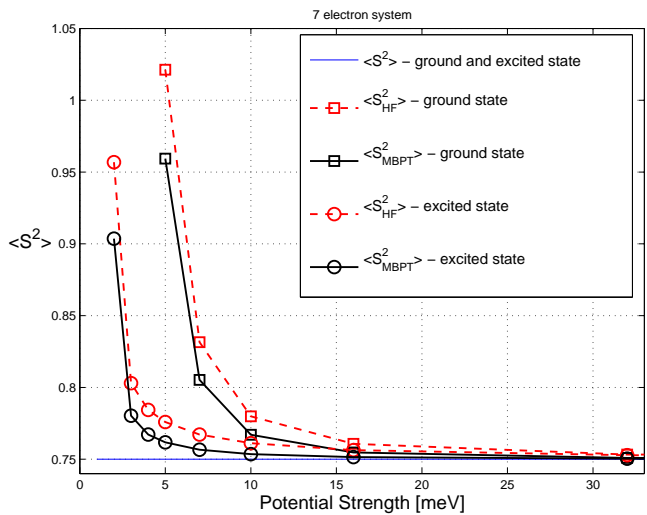}
\caption[]{$\langle S^2 \rangle$ according to Hartree-Fock and second--order
MBPT calculations as functions of the potential strength for the 7
electron ground and excited state.\label{Ssquared7el}}
\end{figure}

Fig.~\ref{Ssquared7el} shows the expectation value of the total
spin, $\langle S^2 \rangle$, according to Hartree-Fock and
second--order MBPT calculations as functions of the potential strength
for the 7 electron ground and excited state. The figure depicts the
drastic onset of spin contamination for weak potentials. While
especially the correlated results, but also the HF--results, converge
towards the correct value for potentials $\ge10$ meV the situation is
worse for weaker potentials. We see that for the ground state the
examined confinement strengths in this article ($\hbar\omega=5$ or $7$
meV) lie on the onset of the spin density wave. It is hard to say how
much this spin contamination affects the energy values but when
compared with the conclusions drawn from Fig. \ref{twoelcompare} and
\ref{sixelcompare}, the energy should not be overestimated with more
than a couple of percent due to spin contamination. For the excited
state the spin contamination is so small (for the $5$ and $7$ meV
calculations) that it should not affect the conclusions from this work.
Moreover we see that, as expected, correlation improves the value of
$\langle S^2 \rangle$.

Table~\ref{tab:Spin} presents the spin contamination for the systems
in the third shell where correlation switched the ground state, namely
the 8, 10 and 11 electron systems. We see that the ground states,
according to our correlated results, are not spin contaminated to any
relevant magnitude. All the excited states are however spin
contaminated. As shown in Fig.~\ref{sixelcompare}, spin contamination
can lower the HF energy and raise the second--order MBPT energy. The
ground state energy switches could thus be an artifact of our starting
point. Energywise however the correlated energies should lie much
closer to the true values than the HF--energies.

\section{Conclusions}
We have shown that the addition of second--order correlation improves
the Hartree-Fock description of two-dimensional few-electron quantum
dots significantly. Our results indicate that details in the addition
energy spectra often attributed to 3D--effects or deviations from
circular symmetry, are indeed sensitive to the detailed description of
electron correlation on more or less the same level. Without precise
knowledge of the many-body effects far reaching conclusions about dot
properties from the addition energy spectra might not be correct.

As a next step we want to include pair-correlation to higher orders to
be able to determine energies with quantitative errors below 0.1meV.
We will then use several different starting potentials to be able to
address also weak confining potentials where the Hartree--Fock
starting point fails.

\begin{acknowledgments}
Financial support from the Swedish Research Council(VR)
and from the G{\"o}ran Gustafsson Foundation is gratefully acknowledged.
\end{acknowledgments}


\end{document}